\documentclass[iop,apj]{emulateapj}
\usepackage{apjfonts}
\usepackage{graphicx,float,wrapfig}
\graphicspath{{Figs/}}
\usepackage{multirow, threeparttable, color}
\usepackage{hyperref}
\hypersetup{
    colorlinks,
    citecolor=blue,
    filecolor=blue,
    linkcolor=blue,
    urlcolor=blue
}

\makeatletter
\newcommand*{\rom}[1]{\expandafter\@slowromancap\romannumeral #1@}
\makeatother 

\newcommand{\fesc}{\ifmmode{f_{\rm esc}}\else{$f_{\rm esc}$}\fi}
\newcommand{\fescs}{\ifmmode{f_{\rm esc}^\star}\else{$f_{\rm esc}^\star$}\fi}
\newcommand{\kms}{\ifmmode{{\;\rm km~s^{-1}}}\else{km~s$^{-1}$}\fi}
\newcommand{\fgas}{\ifmmode{{f_{\rm gas}}}\else{$f_{\rm gas}$}\fi}
\newcommand{\cubecm}{\ifmmode{{\rm cm^{-3}}}\else{cm$^{-3}$}\fi}
\newcommand{\ztwo}{\ifmmode{{\rm [Z_2/H]}}\else{[Z$_2$/H]}\fi}
\newcommand{\zthree}{\ifmmode{{\rm [Z_3/H]}}\else{[Z$_3$/H]}\fi}
\newcommand{\lsim}{\lower0.3em\hbox{$\,\buildrel <\over\sim\,$}}
\newcommand{\gsim}{\lower0.3em\hbox{$\,\buildrel >\over\sim\,$}}

\newcommand{\sfr}{\ifmmode{\textrm{M}_\odot \,\textrm{yr}^{-1} \,\textrm{Mpc}^{-3}}\else{M$_\odot$ yr$^{-1}$ Mpc$^{-3}$}\fi}
\newcommand{\hsfr}{\ifmmode{\;\textrm{M}_\odot\, \textrm{yr}^{-1}}\else{M$_\odot$ yr$^{-1}$}\fi}

\newcommand{\eavg}{\ifmmode{\langle E_\gamma \rangle}\else{$\langle E_\gamma \rangle$}\fi}

\newcommand{\enzo}{{\it Enzo}}
\newcommand{\yt}{{\it yt}}
\newcommand{\moray}{{\it Enzo+Moray}}
\newcommand{\Ms}{\ifmmode{M_\odot}\else{$M_\odot$}\fi}
\newcommand{\vrms}{\ifmmode{v_{\rm rms}}\else{$v_{\rm rms}$}\fi}

\newcommand{\hh}{H$_2$}

\newcommand{\tvir}{\ifmmode{T_{\rm{vir}}}\else{$T_{\rm{vir}}$}\fi}
\newcommand{\mvir}{\ifmmode{M_{\rm{vir}}}\else{$M_{\rm{vir}}$}\fi}
\newcommand{\rvir}{\ifmmode{r_{\rm{vir}}}\else{$r_{\rm{vir}}$}\fi}

\newcommand{\jj}{\ifmmode{J_{21}}\else{$J_{21}$}\fi}
\newcommand{\flw}{\ifmmode{F_{LW}}\else{$F_{LW}$}\fi}
\newcommand{\kph}{\ifmmode{k_{\rm ph}}\else{$k_{\rm ph}$}\fi}

\newcommand{\zsun}{\ifmmode{\rm\,Z_\odot}\else{$\rm\,Z_\odot$}\fi}

\newcommand{\hii}{H {\sc ii}}

\newcommand{\nhi}{\ifmmode{N_{\rm HI}}\else{$N_{\rm HI}$}\fi}
\newcommand\unit[1]{\; \textrm{#1}}

\begin{document}

\shorttitle{Pre-reionization Galaxy Scaling Relations}
\shortauthors{Chen et al.}
\journalinfo{Submitted to the Astrophysical Journal}
\submitted{Draft version \today}

\title{Scaling relations for galaxies prior to reionization}
\author{Pengfei Chen\altaffilmark{1},
  John H. Wise\altaffilmark{2}, 
  Michael L. Norman\altaffilmark{1,3}, 
  Hao Xu\altaffilmark{1},
  and Brian W. O'Shea\altaffilmark{4}}

\affil{$^{1}${CASS,
  University of California, San Diego, 9500 Gilman Drive, La Jolla, CA
  92093; 
  \href{mailto:pec008@ucsd.edu}{pec008@ucsd.edu}, 
  \href{mailto:mlnorman@ucsd.edu}{mlnorman@ucsd.edu},
  \href{mailto:hxu@ucsd.edu}{hxu@ucsd.edu}}}

\affil{$^{2}${Center for Relativistic Astrophysics, School of
  Physics, Georgia Institute of Technology, 837 State Street, Atlanta,
  GA 30332; \href{mailto:jwise@gatech.edu}{jwise@gatech.edu}}}

\affil{$^{3}${SDSC,
  University of California, San Diego, 9500 Gilman Drive, La Jolla, CA
  92093}}

\affil{$^{4}${Lyman Briggs College and Department of Physics and
    Astronomy, Michigan State University, East Lansing, MI 48824;
    \href{mailto:oshea@msu.edu}{oshea@msu.edu}}}


\begin{abstract}

  The first galaxies in the Universe are the building blocks of all
  observed galaxies.  We present scaling relations for galaxies
  forming at redshifts $z \ge 15$ when reionization is just beginning.
  We utilize the ``Rarepeak'' cosmological radiation hydrodynamics
  simulation that captures the complete star formation history in over
  3,300 galaxies, starting with massive Population III stars that form
  in dark matter halos as small as $\sim$$10^6 \Ms$.  We make various
  correlations between the bulk halo quantities, such as virial, gas,
  and stellar masses and metallicities and their respective accretion
  rates, quantifying a variety of properties of the first galaxies up
  to halo masses of $10^9 \Ms$.  Galaxy formation is not solely
  relegated to atomic cooling halos with virial temperatures greater
  than $10^4 \unit{K}$, where we find a dichotomy in galaxy properties
  between halos above and below this critical mass scale.  Halos below
  the atomic cooling limit have a stellar mass -- halo mass
  relationship $\log M_\star \simeq 3.5 + 1.3\log(M_{\rm vir} / 10^7
  \Ms)$.  We find a non-monotonic relationship between metallicity and
  halo mass for the smallest galaxies.  Their initial star formation
  events enrich the interstellar medium and subsequent star formation
  to a median of $10^{-2} \zsun$ and $10^{-1.5} \zsun$, respectively,
  in halos of total mass $10^7 \Ms$ that is then diluted by metal-poor
  inflows, well beyond Population III pre-enrichment levels of
  $10^{-3.5} \zsun$.  The scaling relations presented here can be
  employed in models of reionization, galaxy formation and chemical
  evolution in order to consider these galaxies forming prior to
  reionization.

\end{abstract}

\keywords{galaxies: formation -- galaxies: high-redshift -- methods:
  numerical --- radiative transfer}

\section{Introduction}








Low-mass galaxy observations at $z \ga 6$ naturally have low
signal-to-noise ratios with current telescopes because they are
distant and intrinsically dim.  Nevertheless, recent observational
campaigns have provided valuable constraints on the nature of the
first galaxies and their central black holes (BHs) and their role
during reionization.  The Hubble Ultra Deep Field (HUDF) 2009 and 2012
campaigns \citep{Ellis13} can probe galaxies with stellar masses as
small as $3 \times 10^8 \Ms$ at $z \ga 7$ and as distant as $z \sim
10$ \citep{McLure11, Zheng12, Coe13, Oesch13}.  From the steep slope
of the faint-end of the luminosity function \citep[e.g.][]{Bouwens11,
  Bouwens14_LF, McLure13}, there should be an unseen population of
even fainter and more abundant galaxies that will eventually be
detected by next-generation telescopes, such the \textit{James Webb
  Space Telescope} \citep[JWST, launch date 2018;][]{Gardner06} and
30-meter class ground-based telescopes\footnote{European Extremely
  Large Telescope \citep[E-ELT, 39-m, completion date 2024;][]{ELT},
  Giant Magellan Telescope \citep[GMT, 24.5-m, completion date
  2020;][]{GMT}, Thirty Meter Telescope \citep[TMT, 30-m, completion
  date 2018;][]{TMT}}.

Semi-analytic models and numerical simulations of galaxy formation and
evolution \citep[for respective reviews, see][]{Benson10, Bromm11} are
invaluable tools to connect their photometry, spectra, and imaging to
the physical properties of their stellar population and dynamics and
underlying dark matter (DM) halo.  Before making observational
predictions, it is necessary to correlate galaxy formation with
cosmological structure formation.  There are a few methods to make
this correlation: halo occupation distribution modeling
\citep[e.g.][]{Bullock02}, conditional luminosity function modeling
\citep[e.g.][]{Yang03}, and the abundance matching technique
\citep[e.g.][]{Colin99, Kravtsov99, Conroy06, Guo10}.
\citet[][hereafter BWC13]{Behroozi13} presented a new Markov Chain
Monte Carlo method that utilized the observed star formation rates
(SFRs) and accordingly the stellar masses of galaxies out to $z \simeq
8$ to break the degeneracies suffered in other methods, and they were
able to constrain the specific SFRs (sSFRs) and cosmic SFRs.  Using
this method, they also constrained the intrinsic stellar mass--halo
mass (SMHM) relation to $z = 15$ \citep[BWC13;][]{Behroozi14}.  Their
results are consistent with the observed galaxy stellar mass
functions, sSFRs, and the cosmic SFR over cosmic time in the halo mass
range of $10^{9} - 10^{15} \Ms$.

At the present day, the stellar mass function deviates from the DM
halo mass function at both the low-mass and high-mass extremes
\citep[e.g.][]{Li09_SMF, Bower12}.  The low-mass deficiency can be
attributed to stellar feedback mainly from supernova (SN) explosions
and photo-evaporation from the ultraviolet background
\citep[e.g.][]{Efstathiou92, Bullock00, Gnedin00, Benson02,
  Okamoto08}, while a more efficient feedback mechanism, most likely
arising from active galactic nuclei (AGN), is responsible for the the
high-mass deficiency \citep[e.g.][]{Tabor93, Ciotti01, Croton06,
  Dubois13_AGN2, Genel14, Schaye14}.

The SMHM relation for high-redshift dwarfs does not necessarily have
the same functional form as the present-day one because of different
environmental conditions at high-redshift.  At the high-mass end,
there is some recent observational evidence that the bright-end of the
galaxy luminosity function does not follow a Schechter function, but
possibly a double power-law, at $z \simeq 7$ \citep{Bowler14} that
could indicate that AGN feedback has not yet quenched star formation
in these large galaxies.  At the low-mass end, galaxies could be
forming in a neutral environment, shielded from any ionizing radiation
during reionization, yielding internal stellar feedback as the main
suppressant of star formation \citep{Ricotti08, Salvadori09, Pawlik13,
  Wise14}.

Furthermore, the first galaxies are directly affected by the
radiative, chemical, and mechanical feedback from massive, metal-free
(Population III; Pop III) stars \citep{ABN02, Bromm02_P3, OShea07a,
  2009Sci...325..601T, Greif11_P3Cluster, Hirano13, Susa13, Susa14}.
Radiative and SN feedback can evacuate the majority of the gas from
the host halo \citep{Whalen04, Whalen08_SN, Kitayama04, Kitayama05,
  Alvarez06_IFT, Abel07}, leaving a gas-poor halo that only recovers
by cosmological accretion after tens of Myr \citep{Wise08_Gal,
  Greif10, Wise12_Galaxy, Muratov13, Jeon14}.  The Pop III SNe also
pre-enrich the gas that ultimately assembles the first galaxies to
$10^{-4} - 10^{-3} \zsun$ \citep{Bromm03_SN, Wise08_Gal, Karlsson08,
  Greif10, Wise12_Galaxy}.  Prior to cosmological reionization,
galaxies can form in DM halos as small as $10^7 \Ms$, and these
low-mass ($V_c = \sqrt{GM_{\rm vir}/R_{\rm vir}} < 30 \kms$) galaxies
provide $\sim$40\% of the ionizing photons to reionization, eventually
becoming photo-suppressed as reionization ensues \citep{Wise14}.
Afterward, a small fraction (5--15\%) of the first galaxies survive
until the present day \citep{Gnedin06}, and ultra-faint dwarf galaxies
(UFDs) discovered in the Sloan Digital Sky Survey (SDSS) that surround
the Milky Way could be the fossils of this subset of the first
galaxies.  They are very metal-poor \citep{Kirby08} and are believed
to have had only one or a few early star formation events
\citep{Koch09, Frebel12}. Unlike the Milky Way halo, which was
assembled through multiple merger and accretion events, UFDs likely
did not form via extensive hierarchical merging of bound stellar
systems \citep{Bovill09, Bovill11a, Bovill11b, Salvadori09,
  Simpson13}.

In this paper, we focus on the scaling relations of dwarf galaxies and
their relationship to their halos that assemble during the initial
stages of reionization.  Using cosmological simulations, we extend the
relations found by BWC13 to even higher redshifts and smaller galaxies
that are the building blocks of all observed galaxies.  This work
improves the statistics of the first galaxy properties found by
\citet{Wise14} by a factor of 100, whose simulation only captured the
formation of 32 galaxies by $z \simeq 7$.  In addition to unveiling
the nature of high-redshift dwarf galaxies, our work can provide
valuable constraints on the origin of a subset of dwarf galaxies in
the local Universe.  We have performed a simulation of a survey volume
of 135 comoving Mpc$^3$ that includes a full primordial chemistry
network, radiative cooling from metal species, both Pop III and
metal-enriched star formation and their radiative, mechanical and
chemical feedback.  We first describe our simulation setup in Section
2. Then we present our main results in Section 3. Last we discuss the
findings and possible biases in our simulation in Section 4.

\section{Simulation Setup}

We further analyze the ``Rarepeak'' simulation originally presented in
\citet{Xu13} that focuses on the formation of the first stars and
galaxies in a relatively overdense region with $\langle \delta \rangle
\equiv \langle \rho \rangle / (\Omega_M \rho_c) - 1 \simeq 0.65$ at $z
= 15$ in the entire survey volume of 135 comoving Mpc$^3$.  Here
$\Omega_M$ is the matter density in units of the critical density
$\rho_c = 3H_0^2/8\pi G$.  We perform the simulation with the adaptive
mesh refinement (AMR) cosmological hydrodynamics code \enzo{}
\citep{Enzo}. The adaptive ray tracing module \moray{} is used for the
radiation transport of ionizing photons \citep{Wise11_Moray}, which is
coupled to the hydrodynamics, energy, and chemistry solvers in
\enzo{}.  We have used this simulation to study the number of Pop III
remnants in the first galaxies \citep{Xu13}, their contribution to the
X-ray background \citep{Xu14}, and the imprint of clustered first
galaxies in 21-cm differential brightness temperatures \citep{Ahn14}.
In this paper, we focus on the relationship between the host DM halo
and the stellar and gaseous properties of the galaxies.  A detailed
description of the star formation and feedback models are given in
\citet{Wise12_RP, Wise12_Galaxy} and \citet{Xu13}, and here we give an
overview of the simulation setup and numerical methods.

We generate the initial conditions for the simulation using MUSIC
\citep{Hahn11} at $z=99$ and use the cosmological parameters from the
7-year WMAP $\Lambda$CDM+SZ+LENS best fit \citep{WMAP7}:
$\Omega_M=0.266$, $\Omega_\Lambda=0.734$, $\Omega_b=0.0449$, $h=0.71$,
$\sigma_8=0.81$, and $n=0.963$, where the variables have the usual
definitions.  We use a comoving simulation volume of (40 Mpc)$^3$ that
has a 512$^3$ root grid resolution and three initial nested grids each
with mass resolution eight times higher in each nested grid.  This
corresponds to an effective initial resolution of 4,096$^3$ and DM
mass resolution of $2.9 \times 10^4 \Ms$.  The finest nested grid has
a comoving volume of $5.2 \times 7.0 \times 8.3$ Mpc$^3$ (302
Mpc$^3$).  We allow further refinement in the Lagrangian volume of the
finest nested grid up to a maximum AMR level $l = 12$, resulting in a
maximal spatial resolution of 19 comoving pc.  Refinement is triggered
by either a baryon or DM overdensity of $4 \times \Omega_{\rm
  \{b,DM\}} \rho_c N^{l(1+\phi)}$, respectively.  Here $N = 2$ is the
refinement factor, and $\phi = -0.1$ causes more aggressive refinement
at higher densities, i.e. super-Lagrangian behavior.  We analyze the
simulation at $z = 15$, at which point it has 1.3 billion
computational cells and consumed over 10 million core-hours on the
NICS Kraken and NCSA Blue Waters supercomputers.  The Lagrangian
region at $z=15$ has a comoving volume of $3.8 \times 5.4 \times 6.6$
Mpc$^3$ (135 Mpc$^3$), and we restrict our survey of high-redshift
galaxies to this high resolution region. At this time, the simulation
has a large number ($\sim$1000) of halos with $M$ > 10$^8$ $M_\odot$,
where new formation of Pop \rom{3} stars declines rapidly while the
formation rate of metal-enriched stars continues to increase.  There
are three halos with $M > 10^9 \Ms$ in the refined region at $z = 15$.

Both Pop \rom{3} and metal-enriched stars form in the simulation,
which have distinct formation and feedback models, and we distinguish
them by the total metallicity of the densest star forming cell.  Pop
\rom{3} stars are formed if $[\mathrm{Z/H}] < -4$, and metal-enriched
stars are formed otherwise. We use the same star formation and
feedback models as the ``RP'' simulation in \citet{Wise12_RP} with the
exception of the characteristic mass $M_{\rm char} = 40 \Ms$ of the
Pop III initial mass function (IMF), whereas
\citeauthor{Wise12_Galaxy} considered 100 \Ms.  To select the Pop III
stellar masses, we do not follow the protostellar collapse to high
densities and through their protostellar evolution
\citep[e.g.][]{Susa14}, rather we randomly sample from an IMF with the
functional form
\begin{equation}
  \label{eqn:imf}
  f(\log M) \, dM = M^{-1.3} \exp\left[ -\left(\frac{M_{\rm
          char}}{M}\right)^{1.6} \right] dM
\end{equation}
that behaves as a power-law IMF at $M > M_{\rm char}$ and is
exponentially cutoff below that mass \citep{Chabrier03}.  This choice
of $M_{\rm char}$ is more consistent with the latest results of Pop
\rom{3} formation simulations \citep[e.g.][]{2009Sci...325..601T,
  Greif11_P3Cluster, Hirano13, Susa13, Susa14}.  We treat
metal-enriched star formation with the same prescription as
\citet{Wise09}, which is similar to the Pop \rom{3} prescription but
without the minimum \hh~fraction requirement. This is removed because
the metal-enriched gas can efficiently cool even in the presence of a
strong UV radiation field \citep[e.g.][]{Safranek10}. To ensure that
the stars only form from cold gas, we restrict star formation to gas
with temperatures $T < 1000 \unit{K}$. Unlike Pop \rom{3} star
particles that represent individual stars, metal-enriched star
particles represent a star cluster of some total mass and an assumed
normal (i.e., Kroupa) IMF with minimum and maximum stellar masses
identical to those inferred in the Milky Way. We set the minimum mass
of a star particle to $m_{\star, {\rm min}} = 1000 \Ms$. If the
initial mass does not exceed $m_{\star, {\rm min}}$, the star particle
does not provide any feedback and continues to accrete until it
reaches $m_{\star, {\rm min}}$.


\section{Results}

Our numerical survey focuses on the characteristics and scaling
relations of high-redshift galaxies, and we restrict our analysis to
halos that are resolved by at least 300 DM particles, corresponding to
a mass $M_{\rm vir} \simeq 10^7 \Ms$.  There are 3,338 such halos in
the survey volume at $z=15$.  For these halos, we calculate the virial
radius \rvir~and mass \mvir, using an overdensity $\Delta_{\rm vir} =
178$ relative to the proper critical density
\citep[e.g.][]{2001PhR...349..125B}.  From these halos, we construct
halo merger trees.  We investigate the growth of the most massive
progenitor from each timestep.  We also restrict these progenitor
halos to contribute at least 50\% of their mass to the descendant.
Note that the length of a particular merger tree could be less than
the total number of snapshots.

We study 64 snapshots from $z=18.43$ to $z=15.00$. For all of the halo
quantities in each snapshot, including $M_{\rm vir}$, $M_{\rm gas}$,
$M_\star$ and their accretion rates $\dot{M}_{\rm vir}$, $\dot{M}_{\rm
  gas}$, $\dot{M}_\star$, we use the time difference between that
snapshot and the former snapshot as its weight, since the 64 snapshots
are not equally spaced in time because of computational
reasons\footnote{Snapshots exist at equal time intervals, but we
  utilize the additional outputs that are created at the top-level
  timestep just before the computing queue time limit is reached.}. We
explore the correlations between these properties, plus the stellar
and gas metallicities, by constructing two-dimensional histograms, in
which we include data from all 64 snapshots to increase the sample
size. After the $x$ and $y$ quantities are binned into histograms with
weights, we normalize the histogram values in each $x$-bin with sum
over $y$-axis. Therefore, each histogram value equals the conditional
probability in that $y$-bin given an $x$-value. For each $x$-bin we
calculate the weighted median, 15.9 and 84.1 percentiles, which are
shown as the error bars in all of the Figures.

We fit the weighted medians in most of the phase plots with two
models: (1) a linear model ($\log y =A+\alpha \log x$) and (2) a
smoothed broken power law \citep[SBPL,][]{1999ApL&C..39..281R}. The
logarithmic derivative of a SBPL varies smoothly from $\alpha$ to
$\beta$:
\begin{equation}
  \frac{d\log y}{d\log x}=\xi \tanh\left[\frac{\log
      (x/x_b)}{\delta}\right]+\phi, 
  \label{eqn:dd}
\end{equation}
where $\xi=(\beta-\alpha)/2$, $\phi=(\beta+\alpha)/2$, $x_b$ is the
break point. $\delta$ shows the smoothness of the transition. $\Delta
x=x_b(10^\delta-1)$ gives the linear width of transition. Integrating
Equation (\ref{eqn:dd}) gives
\begin{equation}
  \label{eqn:sbpl}
  \log y=A + \phi \log x + (\ln10) \xi \delta
  \log\left\{\cosh\left[\frac{\log(x/x_b)}{\delta}\right]\right\}.
\end{equation}
Because the SBPL model is insensitive to $\delta$, we fix it to 0.01
in the fits, resulting in four free parameters: \{$A$, $\alpha$,
$\beta$, $x_b$\}.

\begin{table*}
\centering
\caption{Coefficients for the scaling relation fits to the weighted medians}
\label{tab} 
\begin{tabular}{cccccccc}
\hline\hline
Figure & relation  & model & $A$  & $\alpha$ & $\beta$ & $x_b$  & $R^2$\\
\hline                                                                               
\multirow{2}{*}{\ref{m_gas_vs_m_halo}} & \multirow{2}{*}{$M_{\rm gas}$ vs. $M_{\rm vir}$}                                           &linear   & -1.72$\pm$0.37 & 1.10$\pm$0.05 &&& 0.99\\
&&SBPL  & -0.76$\pm$0.33     & 0.72$\pm$0.09      & 1.20$\pm$0.02     & (4.32$\pm$0.78)$\times 10^7$    & 1.00\\
\hline                                                                               
\ref{3_new}(a) & $\dot{M}_{\rm gas}$$/$$\dot{M}_{\rm vir}$ vs. $M_{\rm vir}$   &linear   & -1.95$\pm$0.33 & 0.13$\pm$0.04 &&& 0.71\\
\hline                                                                               
\multirow{2}{*}{\ref{3_new}(b)} & \multirow{2}{*}{$\dot{M}_\star$ vs. $M_{\rm vir}$} 				   &linear   & -13.15$\pm$1.21 & 1.41$\pm$0.15 &&& 0.95\\
&&SBPL  & -13.16$\pm$0.77     & 0.78$\pm$0.15      & 1.97$\pm$0.13     & (1.05$\pm$0.24)$\times 10^8$    & 1.00\\
\hline                                                                               
\multirow{2}{*}{\ref{3_new}(c)} & \multirow{2}{*}{$\dot{M}_\star$$/$$\dot{M}_{\rm vir}$ vs. $M_{\rm vir}$}
&linear   & -5.63$\pm$1.17 & 0.43$\pm$0.14 &&& 0.67\\
&&SBPL  & -5.29$\pm$1.21     & -0.19$\pm$0.24      & 0.88$\pm$0.17     & (9.14$\pm$3.72)$\times 10^7$    & 0.92\\
\hline                                                                               
\multirow{2}{*}{\ref{m_star_vs_m_halo}} & \multirow{2}{*}{$M_\star$ vs. $M_{\rm vir}$}  			
&linear   & -9.88$\pm$0.84 & 1.88$\pm$0.10 &&& 0.99\\
&&SBPL  & -9.05$\pm$0.67     & 1.31$\pm$0.16      & 2.18$\pm$0.08     & (6.97$\pm$1.81)$\times 10^7$    & 1.00\\
\hline                                                                               
\multirow{2}{*}{\ref{2new}(a)} & \multirow{2}{*}{$\dot{M}_\star$ vs. $\dot{M}_{\rm gas}$}                     &linear   & -1.61$\pm$0.15 & 0.50$\pm$0.08 &&& 0.77\\
&&SBPL  & -1.91$\pm$0.23     & 0.28$\pm$0.17      & 0.77$\pm$0.21     & 0.11$\pm$0.21    & 0.82\\
\hline                                                                               
\multirow{2}{*}{\ref{2new}(b)} & \multirow{2}{*}{$\dot{M}_\star$ vs. $\dot{M}_{\rm vir}$}                      &linear   & -2.06$\pm$0.12 & 0.34$\pm$0.08 &&& 0.60\\
&&SBPL  & -2.61$\pm$0.28     & -0.19$\pm$0.28      & 0.55$\pm$0.12     & 0.14$\pm$0.18    & 0.75\\
\hline                                                                               
\ref{Z_vs_m_dm}(a) & [Z/H]$_{\rm gas}$ vs. $M_{\rm vir}$ &SBPL  & -5.79$\pm$1.92     & -0.27$\pm$0.36      & 1.16$\pm$0.31     & (1.02$\pm$0.48)$\times 10^8$    & 0.86\\
\hline                                                                               
\ref{Z_vs_m_dm}(b) & [Z/H]$_\star$ vs. $M_{\rm vir}$ &SBPL  & -4.29$\pm$1.75     & -0.82$\pm$0.33      & 1.32$\pm$0.29     & (1.08$\pm$0.31)$\times 10^8$    & 0.88\\
\hline                                                                               
\ref{Z_vs_m_gas}(a) & [Z/H]$_{\rm gas}$ vs. $M_{\rm gas}$ &SBPL  & -5.10$\pm$1.45     & -0.29$\pm$0.04      & 1.07$\pm$0.37     & (1.01$\pm$0.57)$\times 10^7$    & 0.90\\
\hline                                                                               
\ref{Z_vs_m_gas}(b) & [Z/H]$_\star$ vs. $M_{\rm gas}$ &SBPL  & -8.74$\pm$5.11     & -0.22$\pm$0.06      & 1.99$\pm$1.28     & (4.30$\pm$2.52)$\times 10^7$    & 0.76\\
\hline                                                                               
\ref{Z_vs_m_star}(a) & [Z/H]$_{\rm gas}$ vs. $M_\star$ &SBPL  & -4.27$\pm$0.64     & -0.11$\pm$0.11      & 0.84$\pm$0.18     & (4.17$\pm$2.75)$\times 10^5$    & 0.92\\
\hline                                                                               
\ref{Z_vs_m_star}(b) & [Z/H]$_\star$ vs. $M_\star$ &SBPL  & -4.54$\pm$1.51     & -0.21$\pm$0.17      & 1.01$\pm$0.44     & (7.29$\pm$7.40)$\times 10^5$    & 0.70\\
\hline                                                                               
\multirow{2}{*}{\ref{Z_star_vs_Z_gas}} & \multirow{2}{*}{[Z/H]$_\star$ vs. [Z/H]$_{\rm gas}$}                                         
&SBPL  & -0.37$\pm$0.30     & 0.93$\pm$0.22      & -0.14$\pm$0.32     & 0.10$\pm$0.12    & 0.91\\
&&SBPL$^{\rm a}$ & -0.31$\pm$0.30     & 1.03$\pm$0.10      & -0.32$\pm$0.90     & 0.11$\pm$0.10    & 0.96$^{\rm b}$\\
\hline
\end{tabular}
\parbox[t]{0.73\textwidth}{\textbf{Notes.} 
  All the fits are performed in log-scale for the linear fit, which is
  $\log y = A + \alpha \log x$, and the smoothed broken power law
  (SBPL, Equation \ref{eqn:sbpl}), where $\alpha$ and $\beta$ are the
  slopes below and above the break point $x_b$. The $R^2$ measure
  defined in in Equation (\ref{eqn:rr}) indicates the goodness of the
  fit.  Errors shown have a confidence of 95\%.
  
  $^{\rm a}$ Instead of having the same weight for all the weighted
  medians, this SBPL fit uses the sum of the time weights in each
  x-bin as the weight of the weighted median of data in that x-bin.

  $^{\rm b}$ The definition of $R^2$ for fitting data with weights
  requires changing all the sums in Equation (\ref{eqn:rr}) into
  weighted sums.

}
\end{table*}



All of the fits are performed in log-space, giving a better fit,
because these galactic properties vary over several orders of
magnitude.  The fitting coefficients for all of the relations
presented in this work are shown in Table \ref{tab}.  In the last
column of the Table, we calculate the coefficient of determination
$R^2$ to compare goodness of fit of the two models:
\begin{equation}
 R^2=1-\frac{\sum_i{(\log y_i-f_i)^2}}{\sum_i{(\log y_i-\overline{\log y})^2}},
 \label{eqn:rr}
\end{equation}
where $\overline{\log y}$ is the average of $\log y_i$, and $f_i$ is
the modeled value in log scale. The table shows that the SBPL model
always gives a better fit (i.e. a larger $R^2$) than the linear model
because it has more parameters.

\subsection{Gas accretion rates}

\begin{figure}
\includegraphics[width=\columnwidth]{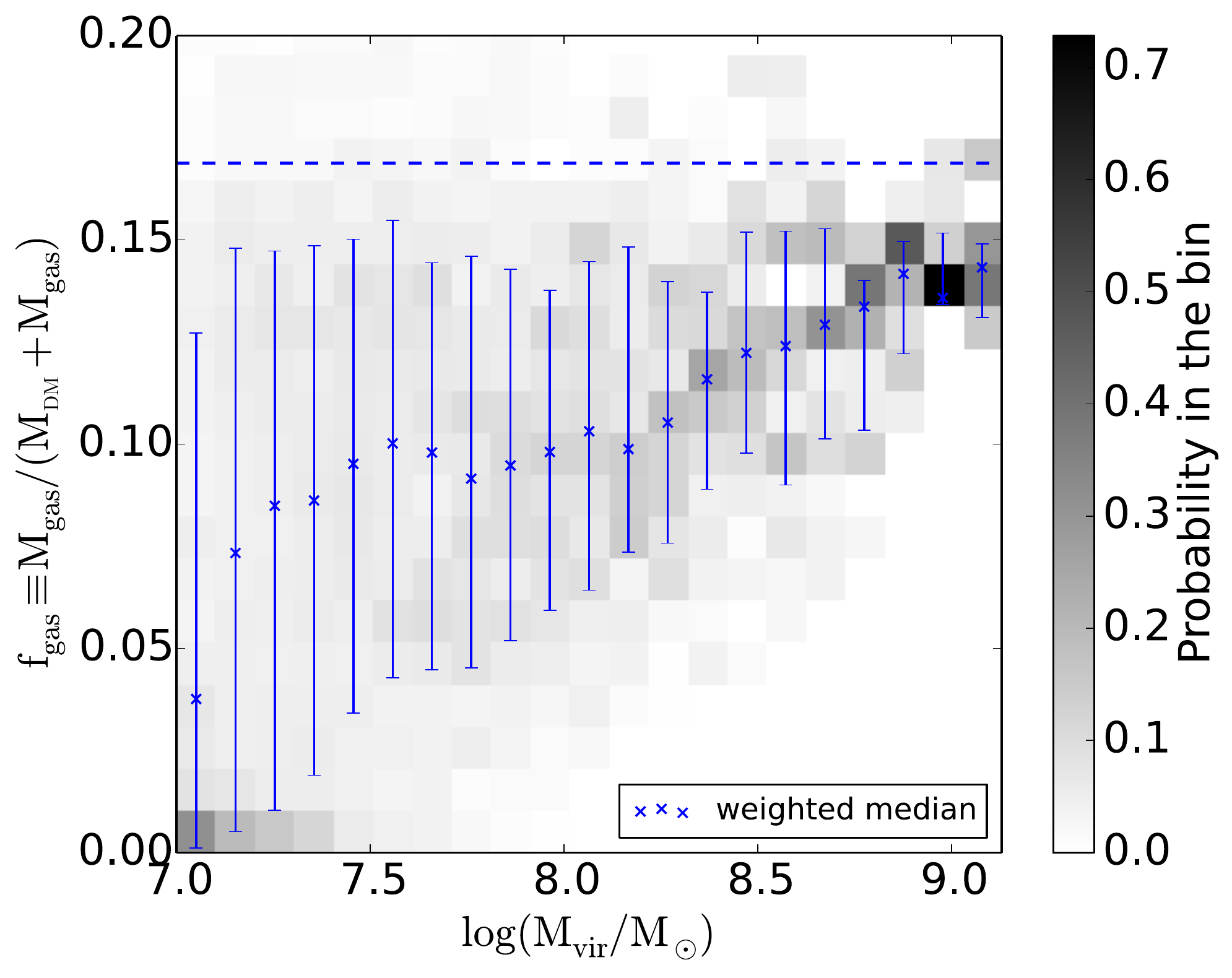}
\caption{Gas fraction of halos as a function of halo mass. As halo
  mass increases, the effect of stellar feedback becomes less severe,
  so gas is accreted more efficiently and the scatter of gas fraction
  decreases.  The dashed horizontal line marks the mean baryon
  fraction $\Omega_{\rm b}/\Omega_{\rm M}$.  The error bars in this
  figure, along with all of the other figures, depict the 15.9 to 84.1
  percentiles in the distribution.}
\label{f_gas_vs_m_halo}
\end{figure}

\begin{figure}
\includegraphics[width=\columnwidth]{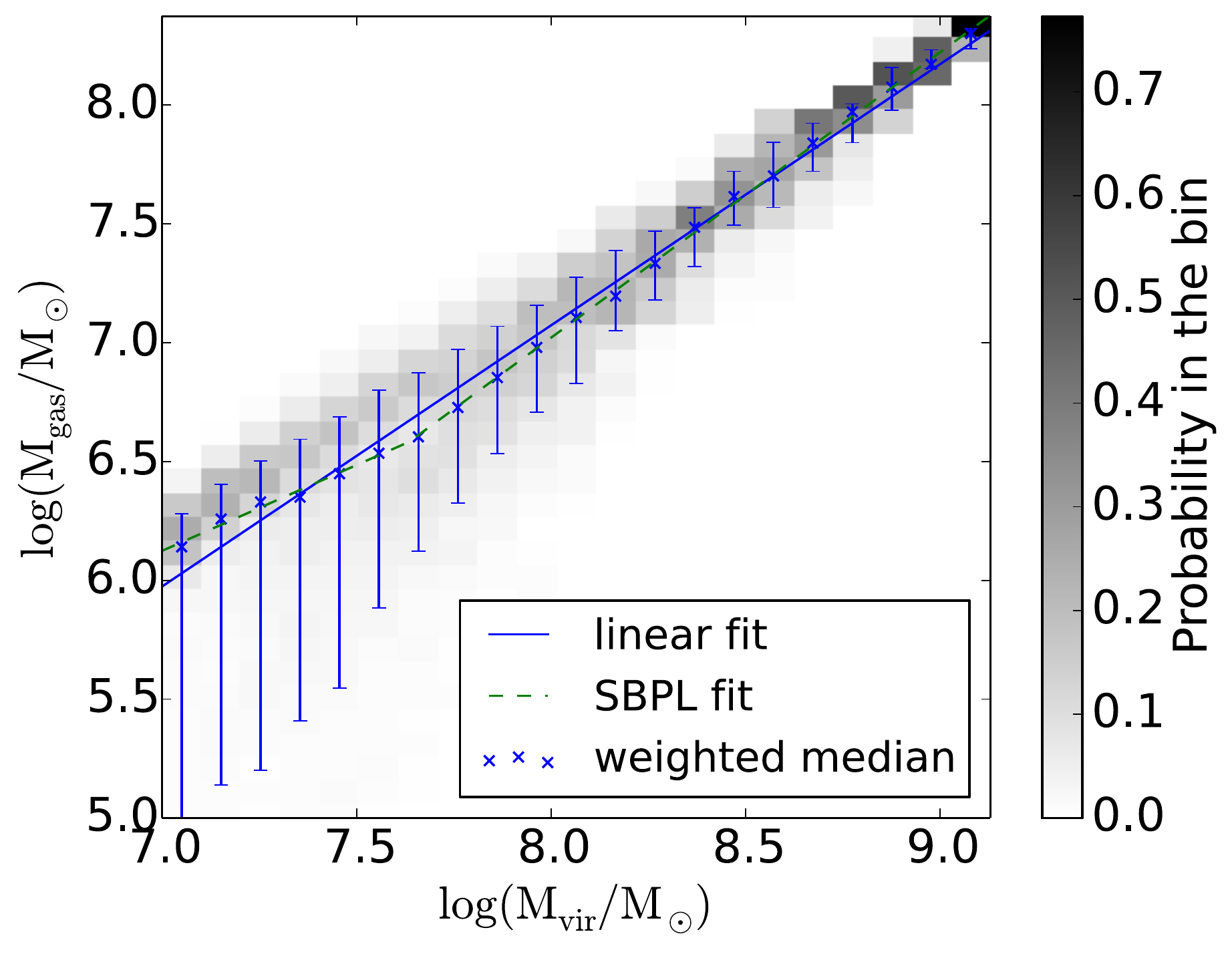}
\caption{Mass of gas in halos as a function of halo mass. With a slope
  of 1.1, mass of gas increases almost linearly with halo
  mass. Fitting results are shown in Table 1.}
\label{m_gas_vs_m_halo}
\end{figure}

Figure \ref{f_gas_vs_m_halo} shows the gas fraction of halos $f_{\rm
  gas}\equiv M_{\rm gas}/(M_{\rm DM}+M_{\rm gas})$ as a function of
halo mass $M_{\rm vir}$. The weighted median of gas fraction increases
with halo mass, while the scatter of gas fraction decreases. At low
masses, halos are susceptible to feedback that is caused either
internally through the shock waves generated by \hii~regions and SNe
or externally through photo-evaporation by a strong ionizing UV flux
originating from nearby galaxies.  The large scatter represents the
different levels of feedback experienced by such halos, for example,
varying Pop III stellar masses and endpoints (i.e. a SN or direct BH
formation) or being embedded in a large-scale neutral or ionized
region.  At $z = 15$, halos with mass $M = 3 \times 10^8 \Ms$ have
circular velocities $V_c \simeq 30 \kms$, holding the gas tenuously in
its gravitational grasp.  Around and above this mass scale, the gas
fractions nearly recover to the cosmic mean gas fraction $\Omega_{\rm
  b} / \Omega_{\rm M}$ as baryons can withstand the effects of stellar
feedback and photo-evaporation.  We show mass of gas as a function of
halo mass in Figure \ref{m_gas_vs_m_halo}, where mass of gas increases
almost linearly with halo mass ($M_{\rm gas} \propto M_{\rm
  vir}^{1.1}$), while the scatter of $M_{\rm gas}$ decreases with
$M_{\rm vir}$.  The slope is greater than unity because there is a
transition from a gas-poor to gas-rich accretion mode in this mass
range.  The SBPL model is physically motivated by the nature of gas
accretion changing in atomic cooling halos, and it provides a better
fit with a break at $(4.32 \pm 0.78) \times 10^7 \Ms$, approximately
the halo mass with a $T_{\rm vir} = 10^4 \unit{K}$ at $z = 15$.  Below
and above the break, the gas mass -- halo mass relation has a slope of
$0.72 \pm 0.09$ and $1.20 \pm 0.02$, respectively.  At higher masses,
we expect the slope to flatten to less than unity so that it never
exceeds \mvir.

\begin{figure}
\includegraphics[width=\columnwidth]{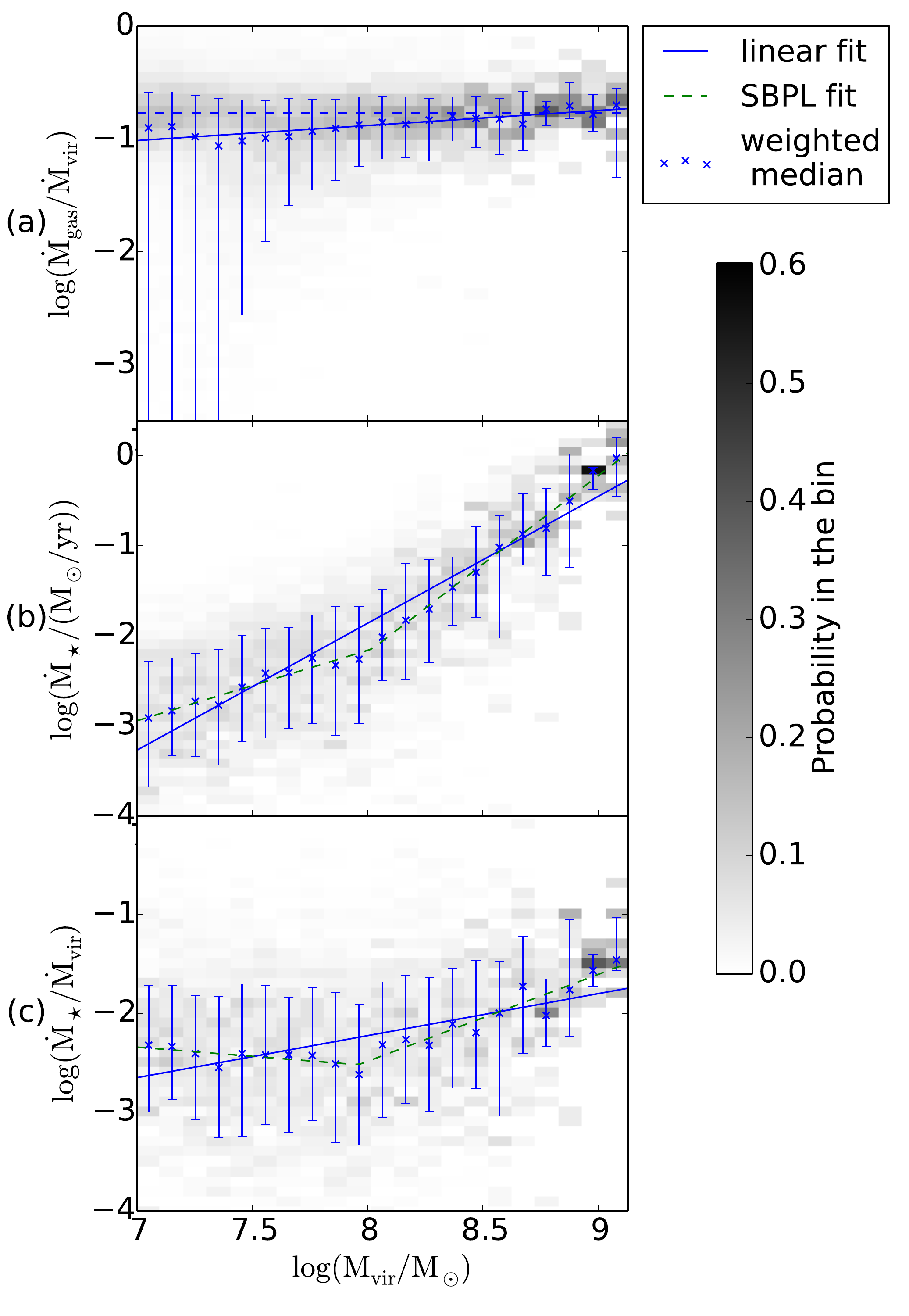}
\caption{(a) Gas accretion rate over halo mass accretion rate as a
  function of halo mass. Negative gas accretion rate is included when
  calculating weighted medians. The weighted medians of this ratio are
  almost constant. Large scatters for low mass halos are due to
  stellar feedback. The lower ends of the first four error bars are
  $-2\times10^{-5}$, $-7\times10^{-5}$, $-5\times10^{-3}$,
  $-9\times10^{-5}$. The dashed horizontal line marks the mean baryon
  fraction $\Omega_{\rm b}/\Omega_{\rm M}$.  (b) Star formation rate
  as a function of halo mass. A clear transition shows up at $M_{\rm
    vir} \sim 10^8 \Ms$, after which the metal-enriched star formation
  becomes more efficient. (c) Star formation rate over halo mass
  accretion rate as a function of halo mass, showing an upturn in star
  formation above $\sim$$10^8 \Ms$.}
\label{3_new}
\end{figure}

Figure \ref{3_new}a shows $\dot{M}_{\rm gas}$$/$$\dot{M}_{\rm vir}$ as
a function of $M_{\rm vir}$. Although $\dot{M}_{\rm
  gas}$$/$$\dot{M}_{\rm vir}$ is almost constant ($\dot{M}_{\rm
  gas}\sim0.1\dot{M}_{\rm vir}$), a small positive slope of the
fitting line means more massive halos are attracting gas more
efficiently.  However in general, halos do not always accrete gas,
where low-mass halos are the most susceptible to strong outflows and
gas loss from radiative and SN feedback because of their shallow
potential wells.  This can be inferred from the observation that
$\sim$13\% of the data points have $\dot{M}_{\rm gas} < 0$, which are
included when calculating weighted medians and curve fittings but not
shown in the histogram.  Most (90\%) of these halos with a net gas
outflow have $M_{\rm vir} < 10^{7.5} M_\odot$.  The large scatter in
the low mass halos are mainly due to variations in $\dot{M}_{\rm
  gas}$. Lower mass halos are more affected by their star formation
history. Specifically, the gas accretion rate depends on the feedback
it has experienced in the past. If the halo hosted a less massive
($\sim 10-30 \Ms$) Pop III star, then less gas would be expelled by
the radiative and SN feedback. However, if the halo hosted a more
massive star ($>60 \Ms$), then most of the gas would have been
expelled due to its higher luminosity.  Massive halos form from the
merger of many smaller halos, which serves to average out the
differences seen in these halos and results in less scatter in the gas
accretion rates.  Another effect that could cause the larger scatter
is environment, that is, whether it is near a larger galaxy and is
partially photo-evaporated by an external radiation field.

\subsection{Star formation efficiency}

The star formation efficiency $f_\star$, i.e. the fraction of gas in a
halo that eventually forms stars, can be influenced by stellar
feedback, internal dynamics, and the size of the gas reservoir
available for star formation, among other factors.  It is an
inherently multi-scale problem, where some fraction of the accreted or
{\it in-situ} gas cools and condenses into dense molecular clouds, and
then some fraction of that molecular gas proceeds to form stars.  Here
we will focus on the connection between star formation and the host
halo and large-scale accretion rates.

\subsubsection{Dependence on halo mass}

\begin{figure}
\includegraphics[width=\columnwidth]{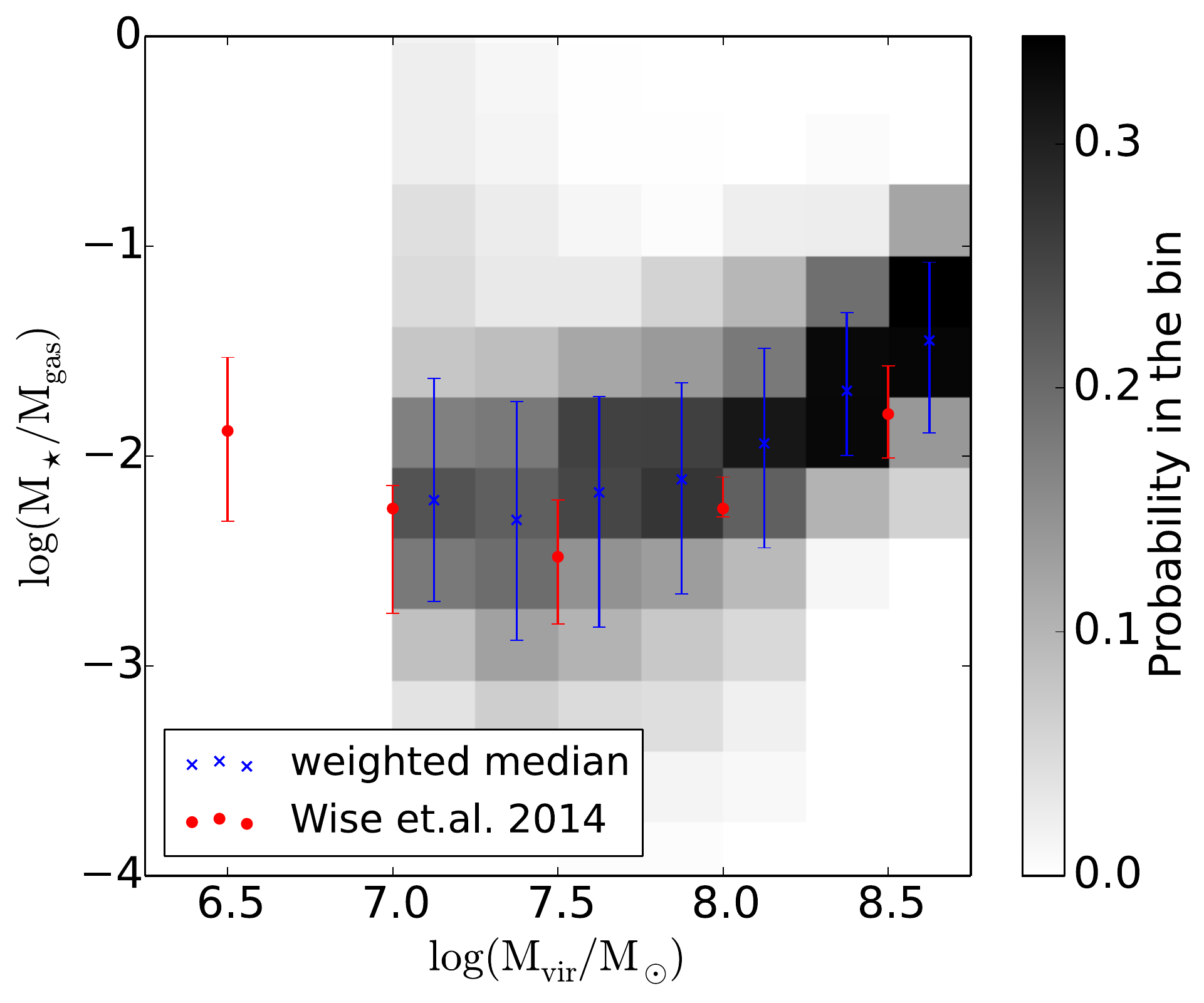}
\caption{Stellar fraction as a function of halo mass.  Each cross
  point is the weighted median of the data points in that x-bin.  For
  halos with $M_{\rm vir} < 10^{7.5} M_\odot$, we restrict the median
  to galaxies with $f_\star < 0.25$ to focus on the star-forming halos.
  These results are consistent with the higher resolution simulation
  of \citet{Wise14}, represented by the red circles.}
\label{f_star_vs_m_halo}
\end{figure}

We show the stellar fraction of halos $f_\star \equiv M_\star / M_{\rm
  gas}$ as a function of halo mass $M_{\rm vir}$ in Figure
\ref{f_star_vs_m_halo}. Most of the low mass halos have $f_\star$
values of around a few percent, because soon after a trace amount of
stars form, \hh~is photo-dissociated throughout the halo by the
radiative, specifically local Lyman-Werner, and SN feedback, first
sterilizing the gas and then disrupting any possible star formation
sites.  This delays the formation of subsequent stars until halos
reach $T_{\rm vir} \sim 10^4 \unit{K}$ \citep{Haiman00, Ciardi00}.
There are 5\% of total halos with a high stellar fractions $f_\star >
0.2$, and nearly all of these halos (98\%) have $M_{\rm vir}$ less
than $10^{7.5} M_{\odot}$.  These high fractions occur when stellar
feedback expels a large fraction of gas, leaving behind a gas-poor
halo, thus increasing $f_\star$.  This behavior was previously shown
in the gas accretion rates, and the gas reservoir only recovers later
through cosmological gas accretion above $T_{\rm vir} = 10^4
\unit{K}$.  In these larger atomic cooling halos, stellar feedback has
less of an effect.  First, the total SN energies are less than the
binding energy of the halo, and, second, the \hii~regions are mostly
contained within the halo, reducing gas blowout from ionization
fronts.  In Figure \ref{f_star_vs_m_halo}, the points show the
weighted medians of all the $f_\star$ at $M_{\rm vir} > 10^{7.5} \Ms$,
and we restrict the median to only include galaxies with $f_\star <
0.25$ below that mass scale to focus on objects that are actively
forming stars in relatively gas-rich halos.

\begin{figure}
\includegraphics[width=\columnwidth]{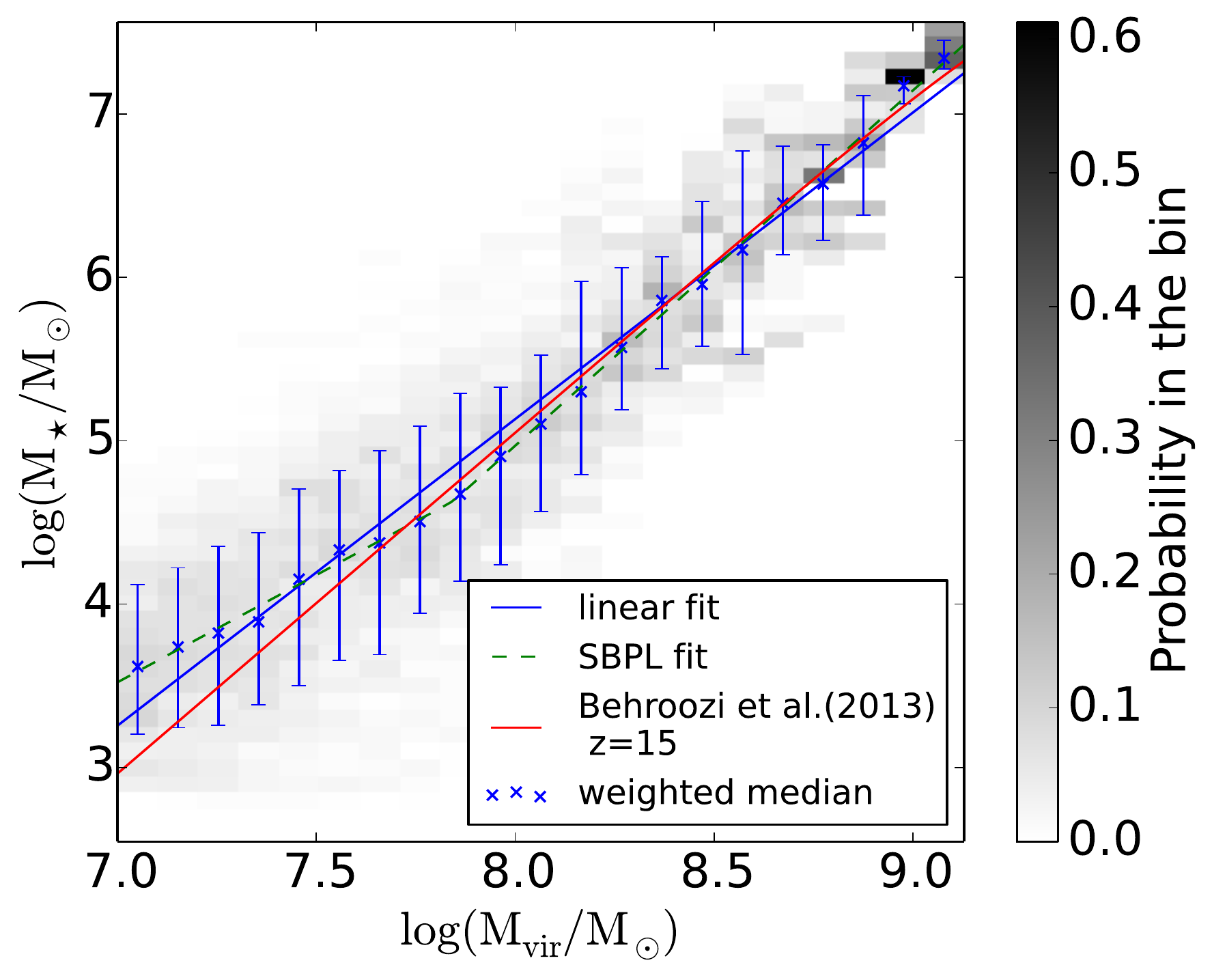}
\caption{The stellar mass--halo mass (SMHM) relation compared with the
  relation from \citet{Behroozi13}.}
\label{m_star_vs_m_halo}
\end{figure}

The SMHM relation is shown in Figure \ref{m_star_vs_m_halo}. Since
there is no observational data at such high redshifts, we compare our
results with the SMHM relation of BWC13 that considers observations up
to $z \simeq 8$.  Since their fitting formula explicitly depends on
redshift and halo mass, we compare their fit at $z = 15$ to our
simulated SMHM relation, even though this extrapolates beyond the
bounds of their original work.  We choose the redshift where the
simulation ends because roughly 60\% of our data points are in the
range $z = 15-16$.  The slope $\alpha = 1.88 \pm 0.10$ of the linear
fit is larger than unity (see Table 1) because this halo mass range
probes a regime where a transition from inefficient to efficient star
formation mode occurs.  As with the gas accretion relations, the SBPL
models this transition better with the slope increasing from $\alpha =
1.31 \pm 0.16$ to $\beta = 2.18 \pm 0.08$ at a break point of $\sim 7
\times 10^7 \Ms$.  The slope $\beta$ is closer to the model of BWC13,
and we expect the slope to flatten at higher masses as star formation
becomes self-regulated after this initial burst during the transition
to an atomic cooling halo.  The model of BWC13 also showed that at $z
\le 6$ the slope becomes flatter at halo masses $M \gtrsim 10^{12}
M_\odot$ at which point galaxies grow primarily through accreting
bound stellar systems instead of forming new stars
\citep[e.g.][]{Conroy09}.  The massive end of the stellar mass
function appears to be approximately in place since $z \sim 1$
\citep[e.g.][]{2006MNRAS.372..537W, 2007ApJ...654..858B,
  2008ApJ...682..919C}.

We show $\dot{M}_\star$ as a function of $M_{\rm vir}$ in Figure
\ref{3_new}b. While the average $\dot{M}_\star$ always increases with
$M_{\rm vir}$, there is a clear change of slope at
$\sim10^8M_\odot$. A similar change could be seen from Figure
\ref{3_new}c, which shows $\dot{M}_\star$$/$$\dot{M}_{\rm vir}$ as a
function of $M_{\rm vir}$. There are several possible contributing
factors to this transition.  First, halos larger than $\sim10^8M_\odot$
are not sensitive to the stellar feedback so they can attract gas more
efficiently. Second, their virial temperature $T_{\rm vir} \ga 10^4
\unit{K}$, at which point hydrogen atomic line cooling becomes
efficient. Lastly, the most massive halos have hosted numerous SN
explosions and have been metal enriched and established a metallicity
floor at lower halo masses.  We explore the correlations between
metallicity and halo properties shortly in Section \ref{sec:metals}.

\subsubsection{Dependence on mass accretion rates}

\begin{figure}
\includegraphics[width=\columnwidth]{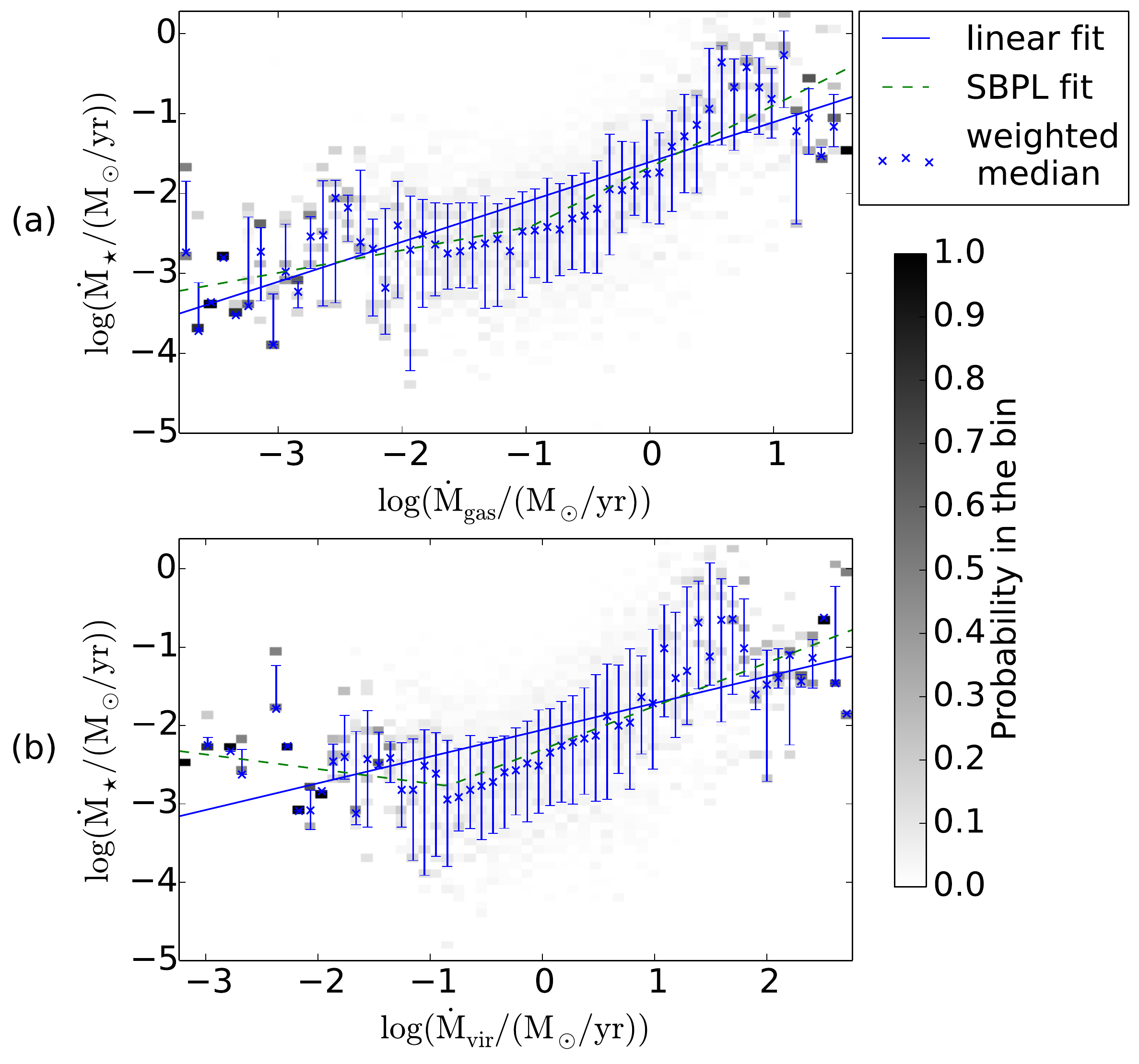}
\caption{(a) Star formation rate as a function of gas mass accretion
  rate. (b) Star formation rate as a function of halo mass accretion
  rate.}
\label{2new}
\end{figure}

We show $\dot{M}_\star$ as a function of $\dot{M}_{\rm gas}$ in Figure
\ref{2new}a, which measures the fraction of accreted gas that forms
stars if we consider this process to be instantaneous.  Overall,
$\dot{M}_\star$ increases as approximately $\dot{M}_{\rm gas}^{0.5}$,
but at $10^{-2.5} \hsfr$ and $10^{0.8} \hsfr$, the increase in stellar
mass is greater than the overall trend but still consistent with the
fit.  We have few DM halos with high ($\gtrsim 10 \hsfr$) or low
($\lesssim 10^{-3} \hsfr$) gas accretion rates, which corresponds to
the high-probability points. Figure \ref{2new}b shows $\dot{M}_\star$
as a function of $\dot{M}_{\rm vir}$.  Above a mass accretion rate
$\dot{M}_{\rm vir} > 0.1 \hsfr$, the slope of this relation is nearly
equal to the $\dot{M}_\star$ -- $\dot{M}_{\rm gas}$ relation because
the gas and mass accretion rates have little dependence on halo mass
above $\sim 10^8 \Ms$ (see Figure \ref{3_new}a).  However below $0.1
\hsfr$, the SFR has little dependence on the mass accretion rates.
This regime occurs when the halo experiences little growth from both
mergers and smooth accretion, so that any increase in stellar mass
must originate from gas that is already present in the halo.  In these
small halos, star formation is regulated to a low SFR of $\sim 10^{-3}
\hsfr$ without any fresh supply of accreted mass to instigate further
star formation.


\subsection{Metallicity}
\label{sec:metals}

The gas-phase and stellar metallicities of galaxies are intimately
related to their SFRs and IMFs.  The first galaxies are both enriched
by Pop III and metal-enriched stars during their assembly.  Here we
investigate the correlations between these metallicities and the halo
mass and growth.  In the following analysis, we calculate the
mass-averaged gas-phase ([Z/H]$_{\rm gas}$) and stellar
([Z/H]$_\star$) metallicities within the virial radius of each halo
with $M_{\rm vir} > 10^7 M_\odot$.  The metallicities include
contributions from SNe originating from both Pop \rom{3} and
metal-enriched stars, and we only include metal-enriched stars in the
stellar metallicity statistics.

\subsubsection{Dependence on virial and gas mass}

\begin{figure}
\includegraphics[width=\columnwidth]{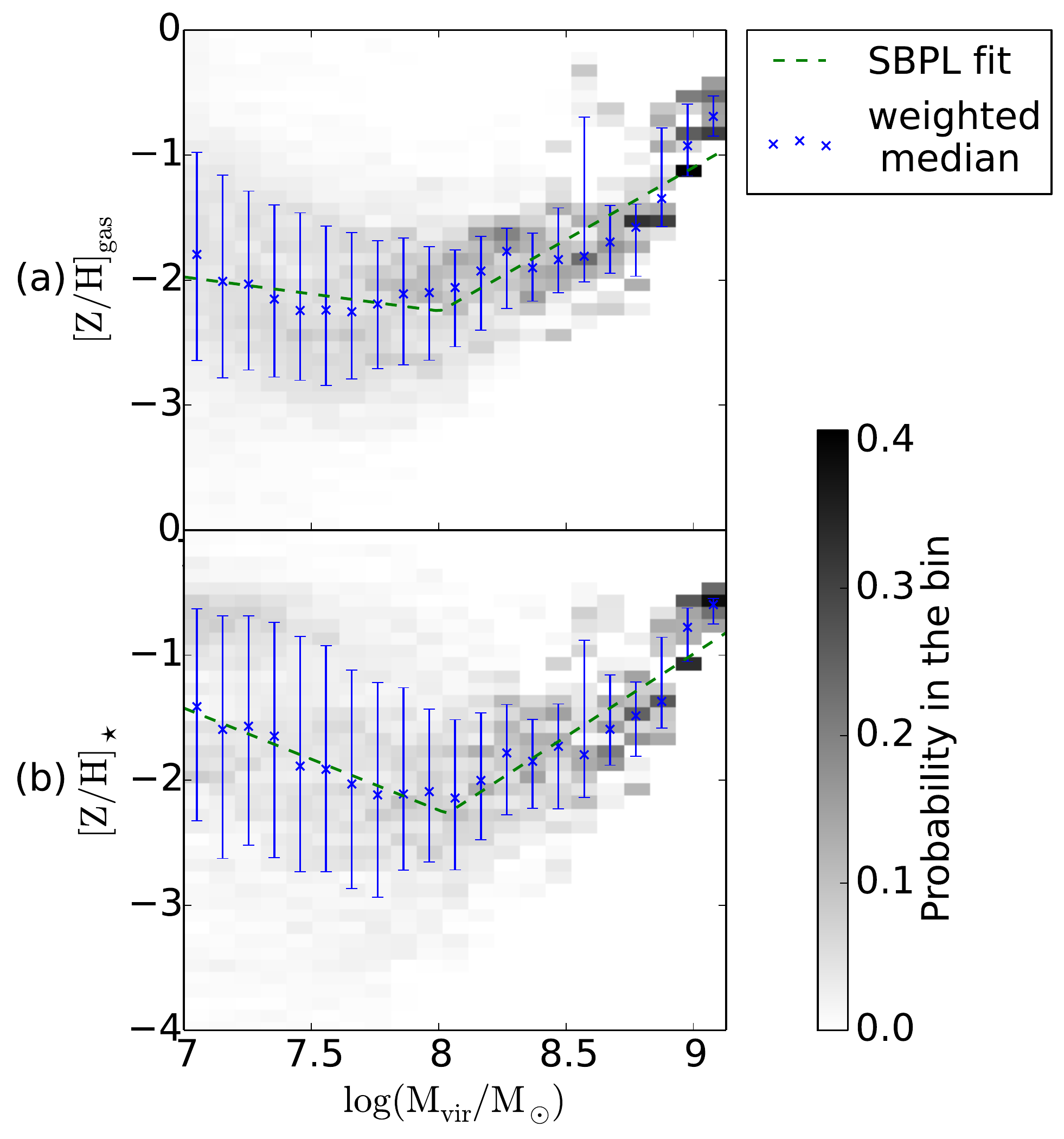}
\caption{Mass-averaged metallicity as a function of halo mass. A
  metallicity floor of [Z/H]$_{\rm gas} \sim -3.5$ is shown for halos
  more massive than $10^{7.5} M_\odot$ with the median metallicity
  1--2 orders of magnitude higher.  After an initial starburst that is
  often triggered by SN blastwaves (biasing the stellar metallicity
  median at the lowest masses), the accretion of pristine and
  metal-poor gas results in the drop of weighted median of gas
  metallicity with increasing halo masses up to $10^8 \Ms$.  In larger
  halos, the metallicity increases because metal-enriched star
  formation becomes efficient, and the metal-rich SN ejecta enrich the
  galaxy.}
\label{Z_vs_m_dm}
\end{figure}

We show both the mean gas metallicity [Z/H]$_{\rm gas}$ and stellar
metallicity [Z/H]$_\star$ as a function of halo mass in Figure
\ref{Z_vs_m_dm}.  The large scatter in [Z/H]$_{\rm gas}$ at low masses
show the dependence on the Pop \rom{3} SN that occurs in the halo
progenitors. Some of the halos with $M_{\rm vir} < 10^{7.5} M_\odot$
have nearly primordial metallicities because they have not hosted a SN
due to the randomness of the Pop \rom{3} IMF. In other words, they
have either formed a Pop \rom{3} star with a BH endpoint or no star
formation has occurred in this halo.  The highest metallicities might
be coming from the Pop \rom{3} SNe exploding in more massive halos,
and the ejecta is being trapped mostly inside of the minihalos
\citep{Whalen08_SN, Ritter12}.  There is a metallicity floor of
[Z/H]$_{\rm gas} \sim -3.5$ for halos more massive than
10$^{7.5}M_\odot$, most of which have been enriched by SN explosions.
Similarly, damped Ly$\alpha$ absorbers (DLAs) have a metallicity floor
of $\sim 10^{-2.8} \zsun$ out to $z \sim 5$
\citep[e.g.][]{Wolfe05_Review, Penprase10, Rafelski12}, and the
metallicity distribution functions of Milky Way halo stars and local
dwarf galaxies precipitately drops below a similar metallicity
\citep[e.g.][]{Battaglia06, Kirby11, McConnachie12, An12}, suggesting
that both extremely metal-poor DLAs and stars have been enriched
primarily by supernovae from Pop III stars \citep[e.g.][]{Bromm03_SN,
  Wise08_Gal, Karlsson08, Greif10, Wise12_Galaxy}.

\begin{figure}
\includegraphics[width=\columnwidth]{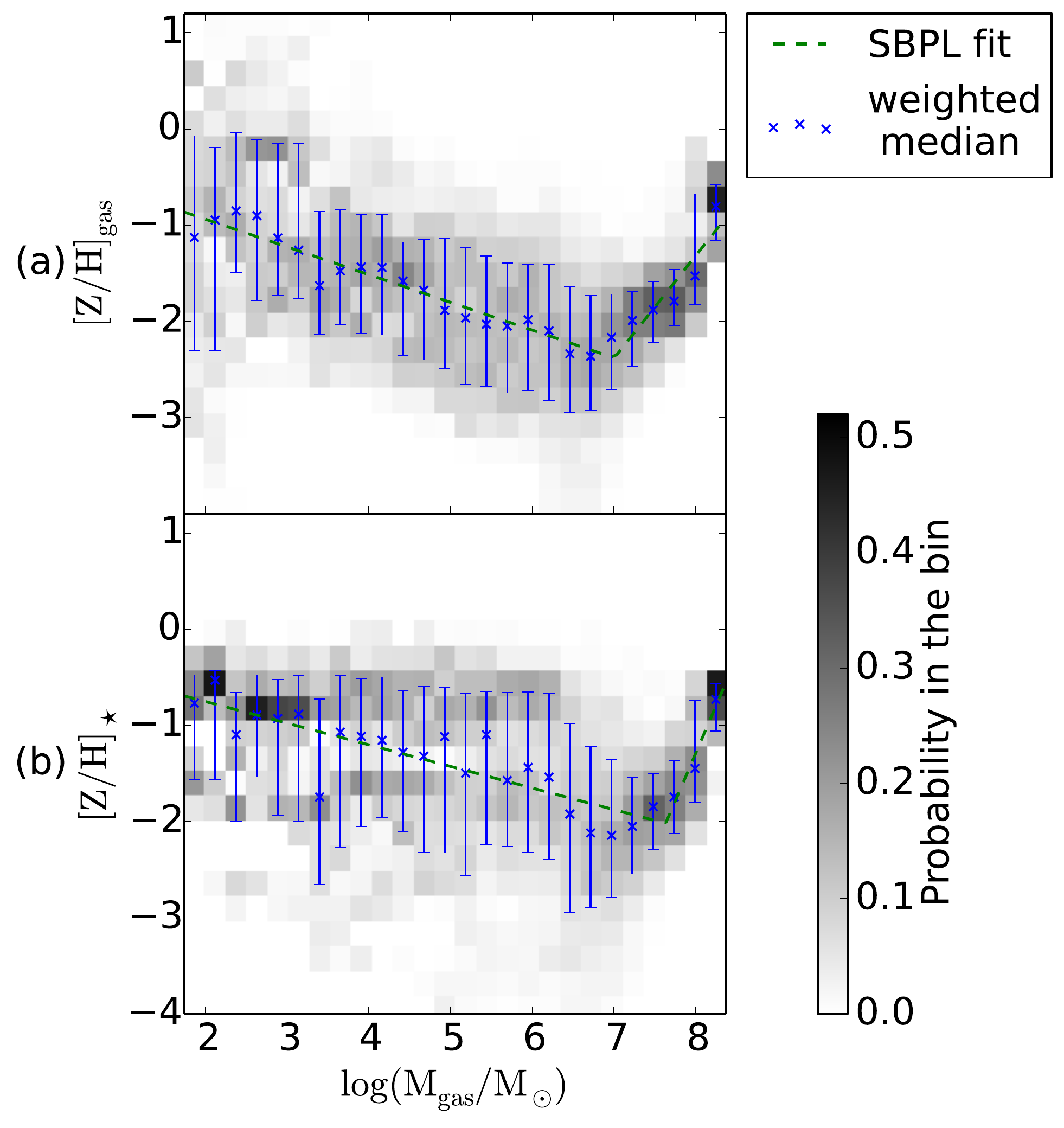}
\caption{Mass-averaged gas-phase (top) and stellar (bottom)
  metallicity as a function of gas mass.}
\label{Z_vs_m_gas}
\end{figure}

We find that the metallicity floor is slightly smaller than
[Z/H]$_{\rm gas} \sim -3$ in \citet{Wise12_Galaxy} because we use a
lower $M_{\rm char}$ in Equation (\ref{eqn:imf}) that favors
hypernovae instead of pair-instability SN.  For instance, the metal
ejecta from a 40 $M_\odot$ hypernova is 8.6 $M_\odot$
\citep{Nomoto06}, compared to 85 $M_\odot$ of metals produced by a 180
$M_\odot$ pair-instability SN \citep{Heger02}, while the hypernova
explosion energies are lower by a factor of a few.  Below a halo mass
of $10^8 \Ms$, the gas-phase metallicity slowly decreases with halo
mass ([Z/H]$_{\rm gas} \propto M_{\rm vir}^{-0.27}$) because the gas
is initially enriched by Pop III SNe and then as metal-free and
metal-poor gas accretes into the halo, this initial enrichment is
diluted.  The effects of dilution is also apparent in Figure
\ref{Z_vs_m_gas} where the gas-phase metallicity is generally lower
than the stellar metallicity.  In larger halos, metal-enriched star
formation becomes more efficient, enriching itself, which is apparent
in the weighted median of [Z/H]$_{\rm gas}$ increasing as $M_{\rm
  vir}^{1.16}$.  Because metal-enriched stars form from this enriched
gas, [Z/H]$_\star$ approximately follows the distribution of
[Z/H]$_{\rm gas}$ in these atomic cooling halos.

However in the lower mass halos, the dilution is more apparent in the
stellar metallicity than the gas-phase metallicity because the star
formation can occur in triggered events as blastwaves overtake nearby
dense clouds inside the halo, leading to high metallicities, which is
reflected in the steeper negative slope in the stellar metallicity
relation.  Afterward, the halo experiences a period without star
formation.  Once the diluted ejecta recollects in the potential well
of the halo, star formation recommences at a lower metallicity,
reflected by the negative slope in the stellar metallicity -- halo
mass relation.  Figure \ref{Z_vs_m_gas} better illustrates this
sequence of enrichment, blowout, and re-accretion in dependence of the
gas-phase and stellar metallicities on the gas mass of the halo.  Here
it is clear that the gas-poor halos have a higher metallicity than
their gas-rich counterparts.  They have most of their gas blown out by
SNe, leaving behind a medium that is more enriched, whereas the more
gas-rich halos have been replenished through further gas accretion.



\subsubsection{Dependence on stellar mass}

\begin{figure}
\includegraphics[width=\columnwidth]{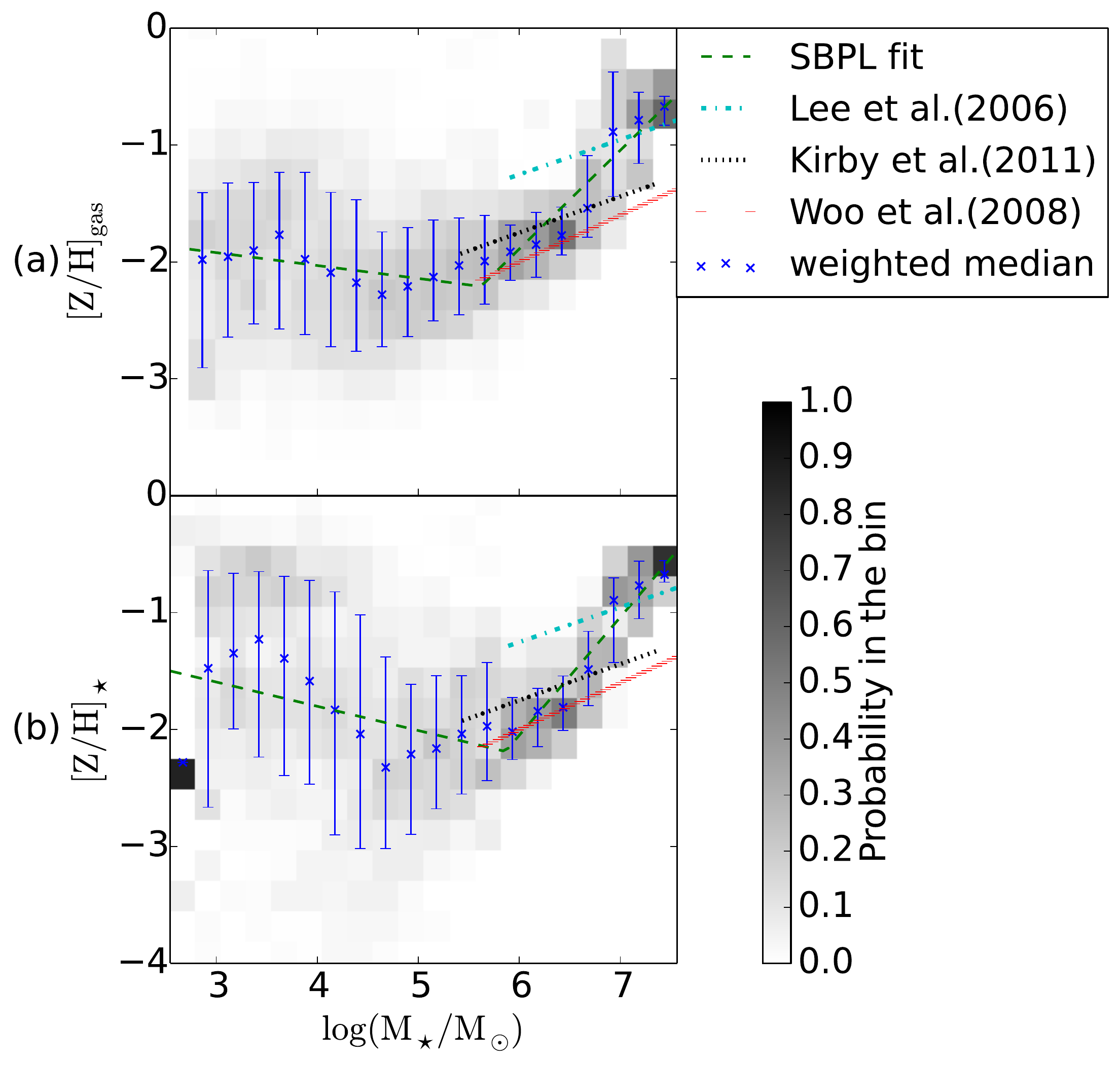}
\caption{Mass-averaged gas-phase (top) and stellar (bottom)
  metallicities as a function of stellar mass compared with three
  observational relations, restricted to their original ranges.}
\label{Z_vs_m_star}
\end{figure}

In Figure \ref{Z_vs_m_star}, we show the metallicity--stellar mass
relations and compare them with a few observational results.  At a
stellar mass $M_\star \la 10^5 \Ms$, the gas-phase and stellar
metallicities decrease as a function of stellar mass for the reasons
described previously.  Above this mass, we can compare our simulated
stellar populations to the metallicity--stellar mass relations in
local dwarf galaxies.  Although this is not a one-to-one comparison,
it gives some constraints on the validity of our simulations because a
fraction of these galaxies will survive until the present-day without
significant star formation \citep[e.g.][]{Gnedin06, Bovill11a,
  Simpson13}.  Furthermore, once the galaxy starts to enrich itself
from {\it in situ} star formation, the stellar metallicity will only
increase from these values at $z=15$.  However, these comparisons
should be viewed as qualitative comparisons because local dwarf
galaxies undergo some amount of tidal harassment as it orbits around
the Milky Way, lowering their stellar content.  In addition, their
metallicity distribution functions will, in principle, be altered
during this process because the more metal-poor stars are less
centrally concentrated than the ``metal-rich'' component in local
dwarfs \citep{Tolstoy04, Battaglia06}.

\citet{Kirby11} used measurements of iron absorption lines in the
central regions of eight dwarf satellite galaxies of the Milky Way,
whose stellar masses range from 10$^{5.4}$ to 10$^{7.4}$
$M_\odot$. \citet{Woo08} analyzed oxygen abundance data of $\sim$40
dwarf galaxies with $M_\star$ ranging from $10^{5.6}$ to $10^{9.6}
M_\odot$. Both of their results fit well with our $\sim$10$^{5.5}$ to
10$^{7}M_\odot$ galaxies, but above this mass range, the metallicities
of the simulated galaxies increase by nearly an order of magnitude.
However, the metallicities in the most massive galaxies in our
simulation better match with the results of \citet{Lee06}, who
analyzed interstellar medium oxygen abundance measurements of 25 dwarf
galaxies with stellar mass $M_\star$ ranging from 10$^{5.9}$ to
10$^{9.3}$ $M_\odot$.  This stellar mass range corresponds to halo
masses $M \ga 10^9 \Ms$ (see Figure \ref{m_star_vs_m_halo}), where we
simulate three halos at this scale.  Because these galaxies are at the
center of large-scale potential well and are experiencing high mass
accretion rates, they may have higher SFRs and thus enrichment rates
than present-day dwarf galaxies, whereas the lower mass galaxies are
possibly regulated by photo-evaporation from a strong UV radiation
field that originates from the central galaxies, keeping their SFRs
and metallicities low.


\subsubsection{Correlation between stellar and gas-phase
  metallicities}

\begin{figure}
\includegraphics[width=\columnwidth]{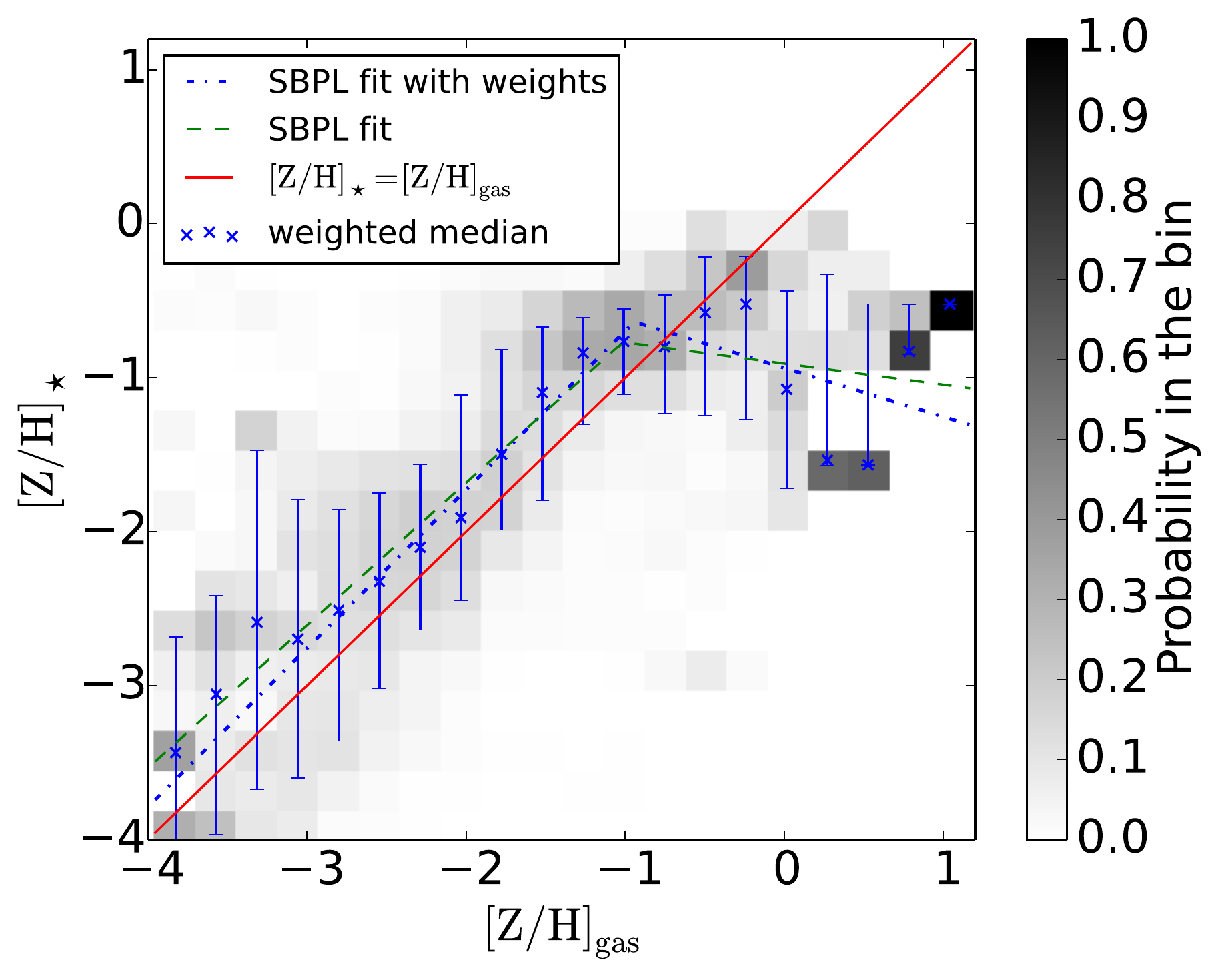}
\caption{Mass-averaged stellar metallicity versus gas metallicity.
  The dash-dotted line shows the SBPL fit to the weighted halo data,
  and the dashed line shows the SBPL fit to the weighted medians.  The
  solid line is the equivalence line for comparison.  Most of the
  halos with low gas metallicity are self-enriched by SN explosions.
  The median of the stellar metallicity is generally higher than the
  gas metallicity, occurring because the SNe ejecta is constantly
  being diluted through a combination of metal-rich outflows and
  metal-poor inflows.}
\label{Z_star_vs_Z_gas}
\end{figure}

Figure \ref{Z_star_vs_Z_gas} compares the correlation between stellar
and gas metallicities.  An equivalence line [Z/H]$_\star$ =
[Z/H]$_{\rm gas}$ is plotted for comparison. For most halos with
[Z/H]$_{\rm gas}<-1$, their stellar metallicities are larger than the
gas-phase metallicities because the gas is constantly being diluted by
a combination of supernova-driven metal-rich outflows and an inflow of
metal-poor gas.  But in most of more metal-enriched halos ([Z/H]$_{\rm
  gas}>-1$), the stellar metallicities are smaller than the gas-phase
ones with 14\% of the data shown in Figure \ref{Z_star_vs_Z_gas} in
this regime, and 85\% of which are in halos of virial mass $M_{\rm
  vir} < 10^8 M_\odot$.  This corresponds to the scenario where
supernova-driven outflow rates are higher than the inflow rates of
metal-poor gas.


\section{Discussion and Conclusions}





In this paper, we present several scaling relations for galaxies that
form in an overdense region at $z \ge 15$, including correlations
between bulk halo properties (virial, gas, stellar masses), their
growth (accretion rates and SFRs), and chemical enrichment (gas-phase
and stellar metallicities).  We derived these relationships from a
zoom-in cosmological radiation hydrodynamics simulation that considers
both metal-free and metal-enriched star formation and feedback.  We
model the effects of radiative feedback using the radiation transport
module \moray.  The high-resolution region contains over 3,300
star-forming halos with the three most massive halos having $M \sim
10^9 \Ms$.

This simulation represents the largest simulated sample of first
galaxies to date and confirms that the first galaxies are not
restricted to atomic cooling halos with virial temperatures $T_{\rm
  vir} \ga 10^4 \unit{K}$, where we have shown that the SMHM extends
down to $10^7 \Ms$ halos, which are adequately sampled by 300 DM
particles in our simulation.  The scaling relations presented here
demonstrate that there is a clear distinction between halos below and
above this critical mass scale corresponding to $T_{\rm vir} = 10^4
\unit{K}$.  Confirming the results of \citet{Salvadori09} and
\citet{Wise14}, halos with $T_{\rm vir} \la 10^4 \unit{K}$ cool
through transitions in \hh{} and metals, forming stars at a slower
rate at $\dot{M}_\star \propto M_{\rm vir}^\alpha$ with $\alpha = 0.78
\pm 0.15$, only to form stars more efficiently at $\alpha = 1.97 \pm
0.13$ when hydrogen line cooling becomes dominant in our sample of
simulated first galaxies.

The correlations presented in this paper, summarized in Table
\ref{tab}, can be utilized in a number of studies that have not
considered star formation in such small halos prior to reionization.
Some examples are (i) semi-analytic models of galaxy formation
\citep[e.g.][]{Benson10, Gomez14}, (ii) reionization calculations
either using semi-analytic, semi-numerical, or $N$-body simulations
\citep[e.g.][]{Zahn11, Alvarez12}, (iii) subgrid models of unresolved
star formation \citep{Trenti07}, and (iv) chemical evolution models
\citep[e.g.][]{Salvadori07, Tumlinson10, Crosby13}.  For instance,
including halos that host low levels of star formation into a
reionization calculation would produce a more extended reionization
history, better matching constraints provided by the cosmic microwave
background \citep{Zahn12, Planck13_Cosmo} and the $z \sim 6$
Gunn-Peterson troughs \citep{Fan06}.  Furthermore, our metallicity
relationships include the pre-enrichment from Pop III stars, so they
provide a robust set of initial conditions for the first galaxies,
evolving to present-day galaxies.

Although our simulation includes most of the relevant physical
processes during the formation of the first galaxies, there are a few
shortcomings in our work that we will address in later studies.
First, the DM resolution only captures the star formation in halos
with masses $M \ge 3 \times 10^6 \Ms$; however in an overdense region,
we expect the intergalactic UV radiation field to be high, with
Lyman-Werner radiation suppressing Pop III star formation in halos
smaller than our resolution limit \citep{Machacek01, Wise07_UVB,
  OShea08}.  Furthermore, supersonic relative velocities between DM
and baryons that originate during recombination can also suppress star
formation in these minihalos \citep{Tselia10, Tselia11, Greif11_Delay,
  Stacy11_Stream, Naoz12}.  We also do not consider the local X-ray
radiative feedback from Pop III BH remnants that could play a role in
further regulating star formation within the galaxy \citep{Alvarez09,
  Xu13, Xu14, Jeon14_XRB}.  Concerning our analysis methods, we
include all of the data in the redshift range $z = 18.4 - 15$ as being
time-independent.  We did this to increase the sample size under the
assumption that the SMHM relation is weakly dependent on redshift, and
furthermore, this redshift range corresponds to only 70 Myr.  We can
also justify this simplification by inspecting the SMHM relation found
in BWC13 that shows the largest difference in stellar mass at a fixed
halo mass is less than 1 dex over all cosmic time.

This work represents an important step forward in quantifying a
variety of properties of the first galaxies (see Table \ref{tab}),
using a sample of over 3,300 star-forming halos at $z \ge 15$.  The
highlights of our findings are as follows.
\begin{enumerate}
\item Halos with virial temperatures $T_{\rm vir} \la 10^4 \unit{K}$
  can cool through \hh{} and fine-structure metal lines, prompting a
  burst of star formation and creating a stellar population with $\log
  M_\star \simeq 3.5 + 1.3\log (M_{\rm vir} / 10^7 \Ms)$.  In atomic
  cooling halos, this slope increases to $\sim$2.2, indicating more
  efficient star formation.  This SMHM relation is consistent with the
  results of \citet{Salvadori09}, BWC13, and \citet{Wise14}.
\item After an initial star formation event that expels most of the
  gas from the halo, the median of the gas fraction never fully
  recovers to the cosmic mean $\Omega_{\rm b}/\Omega_{\rm M}$ in the
  halo mass range presented here.  The median at $10^8 \Ms$ is
  approximately 10\% and increases to 15\% as the halos grow another
  order of magnitude.
\item During this initial star formation event, SN blastwaves further
  enrich the galaxy well beyond Pop III pre-enrichment levels to a
  median of $[{\rm Z/H}] \sim -1.5$ in halos with total and stellar
  masses of $10^7 \Ms$ and $10^{3.5} \Ms$, respectively.  The
  metallicities decrease with halo mass as pristine and metal-poor gas
  accumulates through mergers and smooth accretion.  After the halo
  can cool through atomic line cooling, the galaxy begins to enrich
  itself continuously through sustained and efficient star formation.
\end{enumerate}

Our simulation captures and quantifies the formation of the first
galaxies.  However, we warn that our results are sampled from an
overdense region that is not necessarily representative of the cosmic
mean.  We showed in \citet{Xu13} that the halo mass function at $z =
15$ is about five times that of the cosmic mean and is similar to the
abundances found at $z = 10$.  Our results should not change
significantly in regions of different large-scale overdensities,
although these biased halos could experience higher mass accretion
rates than ones situated in more typical or underdense regions of the
Universe.  To address this issue, we are following up this study with
two more zoom-in simulations of a region of mean matter density, and
also a void region, all within the same (40 Mpc)$^3$ volume, and we
will present their results at a later date.

\acknowledgments

This research was supported by National Science Foundation (NSF) grant
AST-1109243 to MLN.  JHW acknowledges support from NSF grants
AST-1211626 and AST-1333360.  BWO was supported in part by the MSU
Institute for Cyber-Enabled Research and the NSF through grant
PHY-0941373. The simulation was performed on the Kraken supercomputer
operated for the Extreme Science and Engineering Discovery Environment
(XSEDE) by the National Institute for Computational Science, ORNL with
XRAC allocation MCA-TG98020N, and on the Blue Waters operated by the
National Center for Supercomputing Applications (NCSA) with PRAC
allocation support by the NSF (award number OCI-0832662). Data
analysis was performed on the Gordon supercomputer operated for XSEDE
by the San Diego Supercomputer Center and on the Blue Waters
supercomputer.  This research is part of the Blue Waters
sustained-petascale computing project, which is supported by the NSF
(award number ACI-1238993) and the state of Illinois. Blue Waters is a
joint effort of the University of Illinois at Urbana-Champaign and the
NCSA.  This research has made use of NASA's Astrophysics Data System
Bibliographic Services.  Computations and associated analysis
described in this work were performed using the publicly-available
\enzo{} code (\url{http://enzo-project.org}) and the \yt{} toolkit
\citep[\url{http://yt-project.org};][]{yt_full_paper}, which are the
products of collaborative efforts of many independent scientists from
numerous institutions around the world.  Their commitment to open
science has helped make this work possible.

\bibliography{jwise}
\bibliographystyle{apj}

\end{document}